\documentclass[aps,prl,showpacs,superscriptaddress,twocolumn,10pt]{revtex4-1}

\usepackage{graphicx} 	
\usepackage{dcolumn} 	
\usepackage{bm}   		
\usepackage{amssymb}   	
\usepackage[separate-uncertainty=true]{siunitx}		
\usepackage[svgnames]{xcolor}	
\usepackage[version=3]{mhchem}	

\usepackage[caption=false]{subfig}
\usepackage[hidelinks]{hyperref}
\usepackage{cleveref}
\usepackage{lipsum}
\graphicspath{{figs/}}

\newcommand{\uH}{\mu_\mathtt{II}}
\newcommand{\uL}{\mu_\mathtt{I}}
\newcommand{\nH}{n_\mathtt{II}}
\newcommand{\nL}{n_\mathtt{I}}
\newcommand{\pH}{\rho_\mathtt{II}}
\newcommand{\pL}{\rho_\mathtt{I}}

\begin{document}
\title{Enhancement of electron mobility at oxide interfaces induced by \ce{WO3} overlayers}

\author{G. Mattoni\thanks{Contact Address: g.mattoni@tudelft.nl}}
\affiliation{Kavli Institute of Nanoscience, Delft University of Technology, Netherlands}

\author{D. J. Baek}
\affiliation{School of Electrical and Computer Engineering, Cornell University, Ithaca, New York 14853, USA}

\author{N. Manca}
\affiliation{Kavli Institute of Nanoscience, Delft University of Technology, Netherlands}

\author{N. Verhagen}
\affiliation{Kavli Institute of Nanoscience, Delft University of Technology, Netherlands}

\author{L. F. Kourkoutis}
\affiliation{School of Applied and Engineering Physics and Kavli Institute at Cornell for Nanoscale Science, Cornell University, Ithaca, New York 14853, USA}

\author{A. Filippetti}
\affiliation{Dipartimento di Fisica, Università di Cagliari, and CNR-IOM, Istituto Officina dei Materiali, Cittadella Universitaria, Cagliari, Monserrato 09042-I, Italy}

\author{A.D. Caviglia}
\affiliation{Kavli Institute of Nanoscience, Delft University of Technology, Netherlands}

\begin{abstract}

Interfaces between complex oxides constitute a unique playground for 2D electron systems (2DES), where superconductivity and magnetism can arise from combinations of bulk insulators.
The 2DES at the \ce{LaAlO3 / SrTiO3} interface is one of the most studied in this regard, and its origin is determined by both the presence of a polar field in \ce{LaAlO3} and the insurgence of point defects, such as oxygen vacancies and intermixed cations.
These defects usually reside in the conduction channel and are responsible for a decreased electronic mobility.
%
In this work, we use an amorphous \ce{WO3} overlayer to control the defect formation and obtain an increased electron mobility in \ce{WO3 / LaAlO3 / SrTiO3} heterostructures.
The studied system shows a sharp insulator-to-metal transition as a function of both \ce{LaAlO3} and \ce{WO3} layer thickness.
Low-temperature magnetotransport reveals a strong magnetoresistance reaching 900\% at \SI{10}{\tesla} and \SI{1.5}{\kelvin}, the presence of multiple conduction channels with carrier mobility up to \SI{80000}{\centi\metre\squared\per\volt\per\second} and an unusually high effective mass of \SI{5.6}{} $m_\mathrm{e}$.
The amorphous character of the \ce{WO3} overlayer makes this a versatile approach for defect control at oxide interfaces, which could be applied to other heterestrostures disregarding the constraints imposed by crystal symmetry.
\end{abstract}

\date{\today}
\maketitle

The formation of a two-dimensional electron system (2DES) at the interface between band insulators \ce{SrTiO3} (STO) and \ce{LaAlO3} (LAO) is among the most intriguing effects studied in oxide electronics \cite{ohtomo2004high}.
Gate tunable superconductivity \cite{
reyren2007superconducting,
caviglia2008electric
}, strong spin-orbit coupling \cite{
caviglia2010tunable,
diez2015giant
} and magnetism \cite{
bert2011direct,
li2011coexistence
} are some of the many phenomena observed.
The origin of this 2DES is a long standing question in the solid state community and recent results indicate that a consistent picture should take into account both the built-in polar field and the presence of point defects \cite{
nakagawa2006some,
gunkel2012influence,
yu2014polarity
}.
Among these, oxygen vacancies and cation off-stoichiometry in STO are capable of inducing a 2DES \cite{
kalabukhov2007effect,
warusawithana2013laalo3
}.
However, defects residing in the conductive channel are usually responsible for a decreased electronic mobility \cite{	bristowe2011surface
}.
In order to promote high electron mobility, it is crucial to confine donor sites away from the conducting plane, without preventing the 2DES formation in the STO top layers.
Previous attempts to control the defect concentration profile and thus enhance the mobility involved the use of crystalline insulating overlayers \cite{
huijben2013defect,
chen2015extreme
}, adsorbates \cite{
xie2013enhancing
},
amorphous materials \cite{
chen2013high
} and even thin metallic layers \cite{
wu2013nonvolatile,
lesne2014suppression
}.
A promising material to control defect formation is tungsten oxide \ce{WO3}.
The several possible oxidations states of tungsten make \ce{WO3} particularly active in undergoing redox reactions.
For this reason this material is often utilized in electrochemical applications and electrochromic devices \cite{
	deb2008opportunities,
	meng2015electrolyte,
	cong2016tungsten
}.
Also, both crystalline and amorphous \ce{WO3} can host vacancies and interstitial atoms, thus allowing cation accommodation and diffusion, with a tendency to form compounds such as tungsten bronzes \cite{
arab2013strontium,
he2016atomistic
}.
Recent progress demonstrated the high-quality growth of \ce{WO3} thin films on perovskite materials \cite{
	du2014strain,
	leng2015epitaxial,
	altendorf2016facet
}.

In this work we combine the use of a crystalline LAO/STO interface with the high reactivity of amorphous \ce{WO3} to realise a high-mobility metallic 2DES in \ce{WO3 / LAO / STO} heterostructures.
Our approach is based on the tendency of \ce{WO3} to undergo redox reactions, whose contribution is primarily manifested by the reduction of the critical LAO thickness required for the formation of a 2DES.
We characterise the transport properties of this system as a function of \ce{WO3} and \ce{LAO} thickness and find multi-band conduction and an increased electron mobility up to \SI{80000}{\centi\metre\squared\per\volt\per\second}.
The multi-band conduction leads to a remarkably strong classical magnetoresistance, which reaches 900\% at \SI{10}{\tesla} and \SI{1.5}{\kelvin}.
Furthermore, the analysis of Shubnikov-de Haas oscillations unveils an unusually large effective mass of the highly mobile electrons.

\begin{figure}[tb]
\includegraphics[page=1,width=1\linewidth]{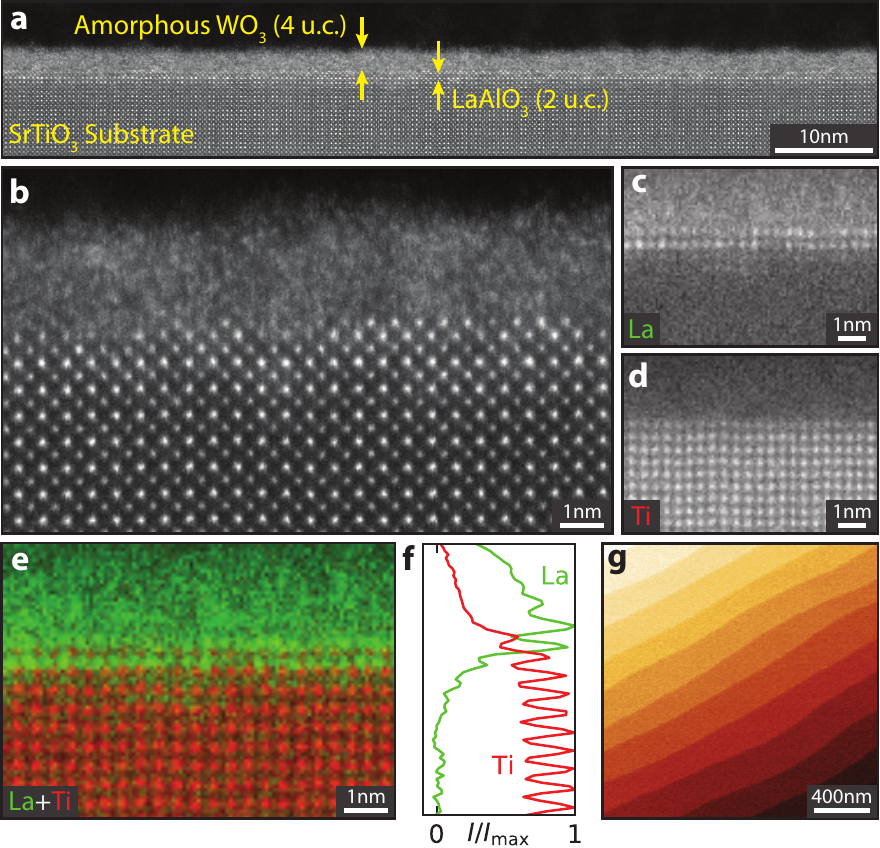}

\subfloat{\label{fig:TEM_large}}
\subfloat{\label{fig:TEM_zoom}}
\subfloat{\label{fig:EELS_Ti}}
\subfloat{\label{fig:EELS_La}}
\subfloat{\label{fig:EELS_Ti_La}}
\subfloat{\label{fig:EELS_profile}}
\subfloat{\label{fig:AFM}}

\caption{\textbf{Structural characterisation of \ce{WO3 / LAO / STO} heterostructures.}
\protect\subref{fig:TEM_large} Lower magnification and 
\protect\subref{fig:TEM_zoom} close-up HAADF-STEM image from a $(4,2)$ heterostructure along the (001) direction.
\protect\subref{fig:EELS_Ti} EELS elemental map showing normalized core-loss signals for La-M$_\mathrm{4,5}$,
\protect\subref{fig:EELS_La} Ti-L$_\mathrm{2,3}$ edges and
\protect\subref{fig:EELS_Ti_La} combined signal.
\protect\subref{fig:EELS_profile} Normalised EELS intensity profile averaged along the direction perpendicular to the interface.
\protect\subref{fig:AFM} Surface topography by AFM.
}

\label{fig:Heterostructures}

\end{figure}

\begin{figure}[tb]
\includegraphics[page=2,width=1\linewidth]{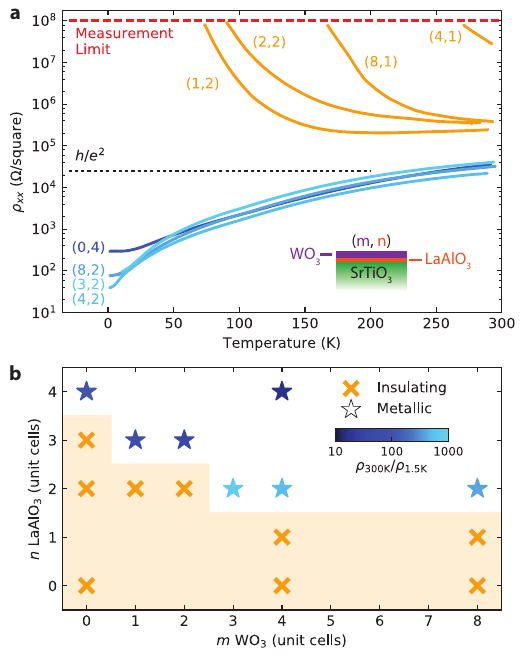}

\subfloat{\label{fig:MIT_thicknesses}}
\subfloat{\label{fig:PhaseDiagram}}

\caption{\textbf{Insulator-to-metal transition in \ce{WO3 / LAO / STO} heterostructures.}
\protect\subref{fig:MIT_thicknesses} Resistance versus temperature for different $(m,n)$ thickness combinations and
\protect\subref{fig:PhaseDiagram} transport phase diagram showing insulating (orange cross) and metallic (stars with colour scale) heterostructures.
The colour scale indicates the residual-resistance ratio for the metallic samples and the orange shaded area marks the $(m,n)$ combinations constituting insulating heterostructures.
}

\label{fig:PhaseDiagram_MIT}

\end{figure}

Ultra-thin heterostructures of amorphous \ce{WO3} and crystalline LAO are grown on \ce{TiO2}-terminated STO (001) substrates by pulsed laser deposition (details on the growth are provided in the supplementary information).
We denote by $(m,n)$ the crystalline equivalent number of unit cells (u.c.) of \ce{WO3} and LAO, respectively, that form the heterostructure.
To investigate structurally the \ce{WO3 / LAO / STO} heterostructure, we perform high-angle annular dark-field scanning transmission electron microscopy (HAADF-STEM).
The HAADF-STEM images in \cref{fig:TEM_large,fig:TEM_zoom} acquired from a $(4,2)$ heterostructure show uniform layers of amorphous \ce{WO3} corresponding to 4 u.c. in thickness, followed by 2 u.c. of crystalline LAO.
Due to the atomic number difference between La and Sr, the HAADF signal from the LAO is more intense than from the underlying STO substrate, as expected.
To further confirm the growth of LAO, electron energy loss spectroscopy (EELS) is subsequently performed.
With an energy dispersion of \SI{0.25}{\electronvolt/channel}, the Ti-L$_\mathrm{2,3}$ and La-M$_\mathrm{4,5}$ edges are recorded simultaneously, providing atomic-resolution Ti and La elemental maps as presented in \cref{fig:EELS_Ti,fig:EELS_La,fig:EELS_Ti_La}.
By averaging the La map parallel to the interface in \cref{fig:EELS_profile}, two clear peaks are shown for La, consistent with the growth of 2 LAO layers in our heterostructure.
However, significant diffusion of La into the \ce{WO3} is also observed.
The surface of all heterostructures is additionally measured by atomic force microscopy (\cref{fig:AFM}), revealing the same regular steps and terraces of the underlying STO substrate, indicating uniform film growth.

The series of resistance versus temperature curves of $(m,n)$ heterostructures in \cref{fig:MIT_thicknesses}, shows a sharp tickness-dependent insulator-to-metal transition. 
The transport measurements are performed in a Van der Pauw configuration (see methods for details).
For different $(m,n)$ combinations the samples show either insulating (orange curves) or metallic (blue curves) character, with a sharp transition between the two regimes as function of layer thickness.
This can be noted comparing the $(4,1)$ and $(4,2)$ curves, where a variation of a single u.c. of the LAO interlayer determines a three orders of magnitude resistivity difference at room temperature, which diverges upon cooling.
It is noteworthy that the onset of the metallic state corresponds to the sheet resistance value $h/e^2$ (dotted line in \cref{fig:MIT_thicknesses}) which is the quantum limit for metallicity in 2D \cite{mott1985minimum}, suggesting this electronic system has a two-dimensional nature.

The interplay between \ce{WO3} and LAO thicknesses is summarised in the phase diagram of \cref{fig:PhaseDiagram}, where we indicate with a shaded orange background the $(m,n)$ combinations resulting in insulating samples.
For LAO-only films $(0,n)$ we reproduce the well-known critical thickness for metallicity of 4 unit cells in crystalline \ce{LAO / STO} interfaces, while samples with only \ce{WO3} $(m,0)$ are always insulating.
Heterostructures with 1 u.c. of LAO $(m,1)$ show an insulating state, independently of the \ce{WO3} layer thickness.
When $n=2$ the insulating state persists for \ce{WO3} thickness $m\leq 2$ only, above which a metallic state is induced.
With 3 cells of LAO a single layer of \ce{WO3} is enough to trigger the metallic state.
We can compare the metallicity of the conducting heterostructures by evaluating their residual resistivity ratio, defined as $\mathrm{RRR}=\rho_{xx}(\SI{300}{\kelvin})/\rho_{xx}(\SI{1.5}{\kelvin})$.
Higher RRR values indicate more pronounced metallic behaviour and are represented by the colour map in \cref{fig:PhaseDiagram}.
Our reference \ce{LAO / STO} heterostructure $(0,4)$ has $\mathrm{RRR}=110$, similarly to previous reports \cite{gariglio2009superconductivity,warusawithana2013laalo3}.
In the \ce{WO3 / LAO / STO} system we find higher values for decreasing thickness of the LAO interlayer.
As an example, the $(4,2)$ combination shows $\mathrm{RRR}= 700$.
A simple interpretation for this trend can be provided by considering two competing effects.
On the one hand the spatially closer the amorphous \ce{WO3} overlayer is to the STO, the more effective it is in controlling defect formation and maintaining a clean conductive channel.
On the other hand a sufficiently thick LAO interlayer is required to provide the polar electric field necessary for driving charge carriers at the \ce{LAO / STO} interface.
The optimal balance of these two effects seems to be achieved for 2 u.c. of LAO, where we measure the highest RRR value.
In this picture we are thus able to combine the mobility enhancement provided by the amorphous overlayer with the advantage of a crystalline conductive interface.

\begin{figure}[b]
\includegraphics[page=3,width=1\linewidth]{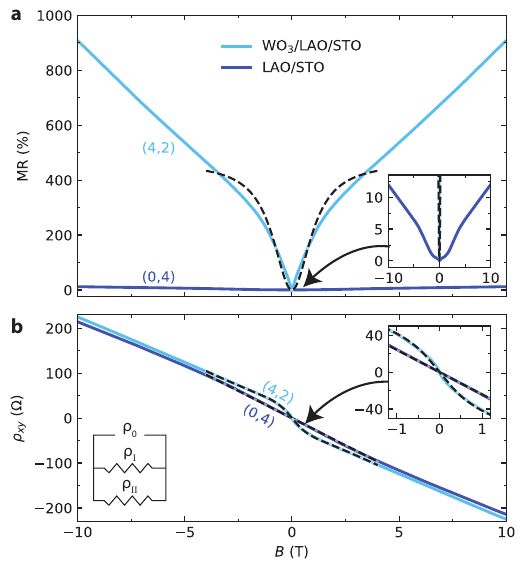}
\subfloat{\label{fig:MR}}
\subfloat{\label{fig:HE}}

\caption{\textbf{Comparison of \ce{WO3 / LAO / STO} and \ce{LAO / STO} magnetotransport.}
\protect\subref{fig:MR} Magnetoresistance and
\protect\subref{fig:HE} Hall effect measured at \SI{1.5}{\kelvin} in Van der Pauw geometry,
with respective zoom-in (insets).
The Hall data is fitted (dashed lines) with \cref{eq:HE_2bands} and used to extract the values in \cref{tab:HE_params}.
The classical MR (dashed line in \protect\subref{fig:MR}) is calculated from these value using \cref{eq:MR_2bands}.
}
\label{fig:MagnetoTransport}

\end{figure}

\begin{table}[b]
\centering
\caption{\textbf{Transport parameters.}
	Mobility, carrier density, sheet resistance and mean free path of the conductive channels extracted from the fits in \cref{fig:HE}.
}
\label{tab:HE_params}

\setlength{\tabcolsep}{3pt}     
\renewcommand{\arraystretch}{1.2} 
\begin{tabular}{l|cccc}
	& $\mu$ (\SI{}{\centi\metre\squared\per\volt\per\second}) & $n_\mathrm{2D}$ (\SI{}{\per\centi\metre\squared}) & $\rho_0$ (\SI{}{\ohm}) & $\lambda$ (\SI{}{\nano\metre}) \\
	\hline
	$(2,4)_\mathtt{II}$ & \SI{80000}{} & \SI{9.3e12}{} & \SI{8}{} & \SI{4100}{} \\
	$(2,4)_\mathtt{I}$ & \SI{3600}{} & \SI{1.7e13}{} & \SI{100}{} & \SI{250}{} \\
	$(4,0)_\mathtt{I}$ & \SI{840}{} & \SI{2.6e13}{} & \SI{290}{} & \SI{70}{}
\end{tabular}

\end{table}
The characteristics of this metallic state are investigated by performing magnetotransport measurements on a $(4,2)$ heterostructure, which shows a high RRR value.
In \cref{fig:MR} we present its magnetoresistance (MR) which is defined as $\mathrm{MR}=\frac{\rho_{xx}(B)-\rho_0}{\rho_0}$, where $\rho_0$ is the sheet resistance at $B=0$ and the magnetic field is applied perpendicular to the interface plane.
At \SI{1.5}{K} the MR is positive and reaches 900\% at \SI{10}{\tesla}, corresponding to one order of magnitude increase in sheet resistance.
This response is very different from what is usually observed in LAO/STO heterostructures $(0,n)$ as can be seen from the comparison with a $(0,4)$ sample in \cref{fig:MR}.
The LAO/STO, in fact, shows a positive MR of only \SI{12}{\percent} at \SI{10}{\tesla}.

The Hall resistance of the $(4,2)$ heterostructure(\cref{fig:HE}) is negative, indicating electronic transport, with a kink at about \SI{1}{\tesla}.
A non-linear component in the Hall effect is typically related to multiple conduction channels contributing to the transport.
In the simplest approximation of two independent channels in parallel, the classical magnetoresistance and the Hall resistance are given by
\begin{subequations}
	\begin{equation}
	\label{eq:MR_2bands}
	\rho_{xx}=
	\frac{(\nL\uL+\nH\uH) + (\nL\uH + \nH\uL)\uL\uH B^2}
	{(\nL\uL+\nH\uH)^2+(n\uL\uH B)^2}
	\cdot\frac{1}{e},
	\end{equation}
	\begin{equation}
	\label{eq:HE_2bands}
	\rho_{xy}=
	\frac{(\pm \nL\uL^2 \pm \nH\uH^2) + n(\uL\uH B)^2}
	{(\nL\uL+\nH\uH)^2+(n\uL\uH B)^2}
	\cdot\frac{B}{e},
	\end{equation}
\end{subequations}
where $n_i$, $\mu_i$ are the carrier density and mobility of the $i$-th channel, $n=(\pm \nL \pm \nH)$ and the $\pm$ sign indicates hole or electron carriers, respectively.
We use \cref{eq:HE_2bands} to fit with good agreement the $(4,2)$ Hall data (dashed line in \cref{fig:HE}) and extract in \cref{tab:HE_params} the corresponding transport parameters (see methods for details).
The $(4,2)$ heterostructure presents two channels of electrons: one with lower-mobility $\uL=\SI{3600}{\centi\metre\squared\per\volt\per\second}$, $\nL=\SI{1.7e13}{\per\centi\metre\squared}$ and one with higher mobility $\uH=\SI{80000}{\centi\metre\squared\per\volt\per\second}$, $\nH = \SI{9.3e12}{\per\centi\metre\squared}$.
Higher mobility values are observed for lower carrier densities, consistent with previous studies of STO-based 2DES \cite{
gunkel2016defect
}.
We note that the sheet resistance of the higher-mobility channel $\pH$ is one order of magnitude smaller than $\pL$, suggesting that it dominates the low-temperature transport.
The $(4,2)$ mobility is about two orders of magnitude higher than what observed in the reference $(0,4)$ sample ($\uL=\SI{840}{\centi\metre\squared\per\volt\per\second}$).
The absence of a higher-mobility channel is coherent with the higher resistivity at \SI{1.5}{\kelvin} and the lower RRR value usually found in \ce{LAO/STO} heterostructures.

Using $n_i$, $\mu_i$ extracted from the Hall effect we calculate with \cref{eq:MR_2bands} the classical two-channels MR (dashed line in \cref{fig:MR}).
The resulting curve accounts for a good extent of the measured signal, which is thus the dominant MR contribution, in particular for small magnetic fields.
The residual MR can arise from the presence of further conduction channels or disorder (Supplementary Figure 5).
Quantum corrections might also be present, but considering their typical magnitude, they are negligible compared to the other contributions.

\begin{figure}[tb]
\includegraphics[page=4,width=1\linewidth]{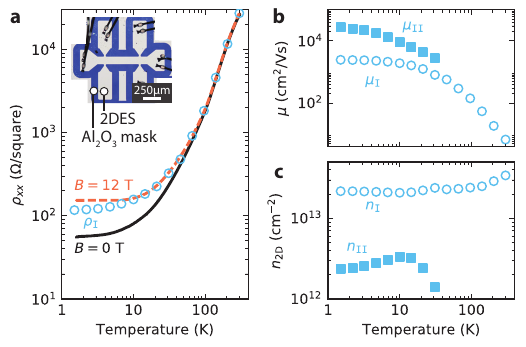}

\subfloat{\label{fig:FieldCool}}
\subfloat{\label{fig:mu_vs_T}}
\subfloat{\label{fig:n_vs_T}}

\caption{\textbf{Temperature dependence of the magnetotransport.}
\protect\subref{fig:MR} Resistance versus temperature with applied perpendicular magnetic field $B=0$ and $B=\SI{12}{\tesla}$ for a $(4,2)$ \ce{WO3 / LAO / STO} heterostructure and resistivity of the lower-mobility channel (circles).
The inset shows an optical image of a \SI{100x500}{\micro\metre} Hall bar used for electrical characterisation.
\protect\subref{fig:mu_vs_T} Mobility and
\protect\subref{fig:n_vs_T} carrier density as a function of temperature for the two channels (squares and circles) extracted from magneto-transport fits with \cref{eq:HE_2bands}.
}

\label{fig:MT_temperature}

\end{figure}

A better insight into the effects of these parallel conduction channels is given by tracking the resistivity and Hall coefficient as a function of temperature.
The measurements are performed in a \SI{100x500}{\micro\metre} Hall bar geometry (inset of \cref{fig:FieldCool}), where the conductive regions are defined using an insulating \ce{Al2O3} hard mask, as described in the methods.
In \cref{fig:FieldCool} we compare the resistivity versus temperature curve measured with a magnetic field of \SI{0}{\tesla} and \SI{12}{\tesla} applied perpendicular to the interface plane.
At \SI{1.5}{\kelvin} the curves are well separated, underscoring a strong positive MR of \SI{200}{\percent}.
On warming, the MR decreases and disappears below our measurement limit around room temperature.

By tracking the Hall effect as a function of temperature (Supplementary Figure 4) we can investigate the temperature dependence of $n_i$, $\mu_i$ for the different channels.
A non-linear Hall effect is observed between \SI{1.5}{\kelvin} and \SI{30}{\kelvin}, while a linear trend is seen at higher temperatures.
The extracted mobilities and carrier densities are presented in \cref{fig:mu_vs_T,fig:n_vs_T}.
In this patterned sample we measure $\uL=\SI{2500}{\centi\metre\squared\per\volt\per\second}$ and $\uH=\SI{27000}{\centi\metre\squared\per\volt\per\second}$ at \SI{1.5}{\kelvin}.
With increasing temperature, at first $\uL$ retains an almost constant value while $\uH$ decreases.
Above \SI{30}{\kelvin} the high-mobility channel disappears and the Hall effect becomes linear, signalling the cross-over to single channel transport.
At higher temperatures $\uL$ decreases several orders of magnitude and reaches $\uL=\SI{7}{\centi\metre\squared/\volt\second}$ at room temperature.
This trend is similar to what has previously been reported for LAO/STO heterostructures \cite{fete2015growth}.

The strong MR in our system can be explained by considering the peculiar characteristics of the two conduction channels.
In general, the classical theory of MR gives a strong resistivity increase with applied magnetic field whenever the charge carriers possess high mobility.
To observe high MR in systems with multiple channels it is also required that the high mobility channel is dominant in the electronic conduction (i.e. $\pH/\pL\ll 1$).
Both conditions are met in our \ce{WO3 / LAO / STO} system, where we find a direct correlation between the ratio $\pH/\pL$ and the MR magnitude at \SI{10}{\tesla}: with $\pH/\pL\sim \SI{e-1}{}$ in \cref{fig:MagnetoTransport} we measure $\mathrm{MR} \sim 900\%$, and with $\pH/\pL\sim 1$ in \cref{fig:MT_temperature} we have a lower $\mathrm{MR} \sim 200\%$.
A further confirmation of this behaviour is given by considering that $\pL$ values in \cref{fig:FieldCool} well represent the resistivity versus temperature curve at $B=\SI{12}{\tesla}$.
This indicates that the high-mobility channel is suppressed in the transport at high magnetic field.

\begin{figure*}[tb]
\includegraphics[page=5,width=1\linewidth]{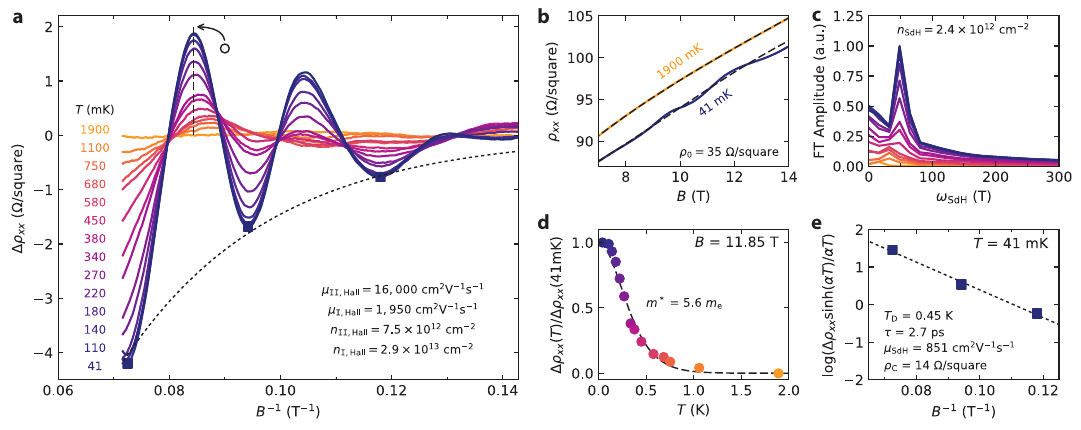}

\subfloat{\label{fig:SdH_dRxx}}
\subfloat{\label{fig:SdH_Rxx}}
\subfloat{\label{fig:SdH_FFT}}
\subfloat{\label{fig:SdH_ampl}}
\subfloat{\label{fig:SdH_dingle}}

\caption{\textbf{Quantum oscillations of conductance.}
\protect\subref{fig:SdH_dRxx} Temperature dependence of the SdH oscillations after the removal of a 3\textsuperscript{rd} order polynomial background.
The dashed and dotted lines indicate the data in \protect\subref{fig:SdH_ampl} and \protect\subref{fig:SdH_dingle}, respectively.
\protect\subref{fig:SdH_Rxx} Raw $\rho_\mathrm{xx}$ data at two different temperatures showing the fit of the polynomial background (dashed lines).
\protect\subref{fig:SdH_FFT} Fourier spectra of the oscillations in the range \SIrange{7}{14}{\tesla}.
\protect\subref{fig:SdH_ampl} Temperature dependence of the oscillations amplitude at $B=\SI{11.85}{\tesla}$ and fit to \cref{eq:SdH} (dashed line) from which the carrier effective mass $m^*$ is extracted.
\protect\subref{fig:SdH_dingle} Dingle plot of the SdH oscillations minima at $T=\SI{41}{\milli\kelvin}$ and fit to \cref{eq:SdH} (dotted line) from which the Dingle temperature $T_\mathrm{D}$, elastic scattering time $\tau$, mobility $\mu_\mathrm{SdH}$ and the classical sheet resistance $\rho_\mathrm{c}$ are extracted.
}

\label{fig:SdH}

\end{figure*}

The carrier density of the two conduction channels present opposite trends as a function of temperature.
At \SI{1.5}{\kelvin} we find that the lower-mobility channel has a higher density $\nL=\SI{2.2e13}{\per\centi\metre\squared}$, and the higher-mobility has a lower-density $\nH=\SI{2.4e12}{\per\centi\metre\squared}$.
Upon warming, $\nL$ maintains an almost constant value, while $\nH$ undergoes a sharp drop above \SI{10}{\kelvin} and subsequently disappears.
This disappearance might be due to the activation of interband scattering processes at higher temperatures, which cause a mixing of $\pL$, $\pH$, so that their populations cannot be independently resolved in Hall effect measurements \cite{gunkel2016defect}.
Another possible interpretation for this trend is that the two conduction channels are situated in STO at two different distances from the interface.
The first channel might be spatially closer to the LAO layer, where electrons experience more defects and a stronger polar electric field, resulting in lower mobility and higher carrier density.
The second channel, instead, could be further away from the interface, where a less-defected STO determines a higher electron mobility.
In this picture, the depopulation of $\pH$ might be linked to the drop of the STO dielectric constant upon warming \cite{
	sakudo1971dielectric
}
(Supplementary Fig. 6).


The electronic state confined in our \ce{WO3 / LAO / STO} heterostructures shows Shubnikov-de Haas (SdH) oscillations superimposed on the background of strong positive MR.
The SdH as a function of temperature are shown in \cref{fig:SdH_dRxx}, where their signal was extracted by fitting the background with a 3\textsuperscript{rd} order polynomial (dashed line in \cref{fig:SdH_Rxx}).
The oscillations disappear when the magnetic field is applied parallel to the interface plane, as expected for a two-dimensional system.
SdH oscillations in 2DES can be modelled by
\begin{equation}
\label{eq:SdH}
\Delta \rho_{xx}=
4\rho_\mathrm{c}\mathrm{e}^{-\alpha T_\mathrm{D}}\frac{\alpha T}{\sinh(\alpha T)}\sin\left(2\pi \frac{\omega_\mathrm{SdH}}{B}\right),
\end{equation}
where
$\rho_\mathrm{c}$ is the classical sheet resistance in zero magnetic field,
$\alpha=2\pi^2 k_\mathrm{B}/\hbar \omega_\mathrm{c}$ with
cyclotron frequency $\omega_\mathrm{c}=eB/m^*$,
Boltzmann's constant $k_B$,
reduced Planck's constant $\hbar$,
carrier effective mass $m^*$ and
Dingle temperature $T_D$.
Fourier analysis in \cref{fig:SdH_FFT} reveals that the oscillations are periodic in $B^{-1}$, with a single frequency peak at $\omega_\mathrm{SdH}=\SI{49}{\tesla}$.
Assuming a 2DES with circular sections of the Fermi surface, we can estimate the carrier density as $n_\mathrm{SdH}=\omega_\mathrm{SdH}\nu_\mathrm{s}e/h$, where $\nu_\mathrm{s}$ indicates the spin degeneracy.
By considering $\nu_\mathrm{s}=2$ we find $n_\mathrm{SdH}=\SI{2.4e12}{\per\centi\metre\squared}$.
We note that even if Hall effect measurements indicate the presence of two conduction channels (values in \cref{fig:SdH_dRxx}), only one channel contributes to the quantum oscillations.
Furthermore, the obtained $n_\mathrm{SdH}$ is lower than both $\nH$, $\nL$ for this sample, so that it is not possible to associate the SdH oscillation to one specific channel.
A discrepancy between $n_\mathrm{SdH}$ and $n_\mathrm{Hall}$ in \ce{LAO/STO} interfaces has already been reported and its origin remains unknown \cite{
	caviglia2010two,
	shalom2010shubnikov
}.

To extract the mass of the electrons showing the SdH effect, in \cref{fig:SdH_ampl} we track the oscillation amplitude at $B=\SI{11.85}{\tesla}$ as a function of temperature (similar results are obtained using different values of $B$).
Fitting the trend with \cref{eq:SdH}, we find a surprisingly high value $m^*=5.6\, m_\mathrm{e}$.
Considering the enhanced mobility of carriers in the \ce{WO3 / LAO / STO} system, in fact, one would expect a decreased effective mass, while in this case $m^*$ is three times larger than typical observations in \ce{LAO / STO} heterostructures \cite{
chen2013high,
mccollam2014quantum
}.

A possible explanation of this electron mass renormalization can be lead back to strong electron-phonon coupling, which is enhanced by the tight spatial confinement of the 2DEG.
Such coupling was previously found to produce large phonon-drag
\cite{
pallecchi2015giant,
pallecchi2016large
}
and polaronic effects in both \ce{LAO / STO} interfaces and amorphous \ce{WO3} thin films
\cite{cancellieri2016polaronic,
berggren2001polaron
}.
Another possibility is that the modified defect profile with respect to conventional \ce{LAO / STO} interfaces determines a mass enhancement of the 2DES bands \cite{wunderlich2009enhanced}.

Finally, from the Dingle plot in \cref{fig:SdH_dingle} we extract $T_D=\SI{.45}{\kelvin}$.
This value points to an ordered electronic system with sharp Landau levels, considering that their energy smearing $k_\mathrm{B}T_\mathrm{D}\sim\SI{40}{\micro\electronvolt}$ is much smaller than their spacing $\hbar\omega_\mathrm{C}\sim\SI{250}{\micro\electronvolt}$.
The extracted value $\rho_\mathrm{C}=\SI{14}{\ohm/square}$ is in good agreement with $\rho_0=\SI{35}{\ohm/square}$, corroborating the performed analysis.
Using $T_D = \hbar / 2\pi k_\mathrm{B}\tau$ and $\tau = m^*\mu_\mathrm{SdH}/e$ we calculate the elastic scattering time $\tau=\SI{2.7}{\pico\second}$ and the quantum mobility $\mu_\mathrm{SdH}=\SI{851}{\centi\metre\squared\per\volt\per\second}$.
Even though $\mu_\mathrm{SdH}$ is lower than both the Hall effect values $\uH$, $\uL$, it further confirms the formation of a high mobility 2DES in the \ce{WO3 / LAO / STO} heterostructure.

To conclude, we have demonstrated that amorphous \ce{WO3} is an effective overlayer to form 2DES with enhanced mobility and effective mass at \ce{LAO / STO} interfaces.
Reducing the crystalline \ce{LAO} critical thickness from 4 to 2 unit cells, the \ce{WO3} overlayer determined a metallic system with high RRR and increased electron mobility.
We ascribed the insurgence of a strong classical magnetoresistance to the peculiar characteristics of the multiple conduction channels observed in the system.
Quantum oscillations of conductance confirmed the realisation of high-quality \ce{WO3 / LAO / STO} heterostructures, where a strong two-dimensional confinement of carriers is achieved.
All these results are achieved using an amorphous \ce{WO3} overlayer, which does not require crystal matching.
Our work thus demonstrates a new approach for defect control at oxide interfaces, which can be exploited to induce high-mobility 2DES in a broad variety of oxide materials.

\section{Experimental Section}
{\small
\textit{Samples growth:}
\ce{WO3 / LaAlO3 / SrTiO3} heterostructures were grown by pulsed laser deposition on commercially available \SI{5x5}{\milli\metre} \ce{SrTiO3} (001) substrates, with \ce{TiO2} surface termination.
The laser ablation was performed using a KrF excimer laser (Coherent COMPexPro 205, $\lambda=\SI{248}{\nano\metre}$) with a \SI{1}{\hertz} repetition rate and \SI{1}{\joule\per\centi\metre\squared} fluence.
The target-substrate distance was fixed at \SI{55}{\milli\metre}.
For the \ce{LaAlO3} thin films a crystalline target was employed and the deposition performed at \SI{800}{\celsius} substrate temperature and \SI{3e-5}{\milli\bar} oxygen pressure.
\ce{LaAlO3} film thickness was monitored \textit{in-situ} during growth by intensity oscillations of reflection high-energy electron diffraction (RHEED).
The samples were annealed for \SI{1}{\hour} at \SI{600}{\celsius} in \SI{300}{\milli\bar} of \ce{O2} atmosphere to compensate for the possible formation of oxygen vacancies.
The amorphous \ce{WO3} thin films were deposited from a \ce{WO3} sintered target at \SI{500}{\celsius} substrate temperature and \SI{5e-3}{\milli\bar} oxygen pressure.
\ce{WO3} film thickness was calibrated by depositing crystalline \ce{WO3} on \ce{SrTiO3} and monitoring the growth by RHEED.
The thickness value was then confirmed by X-ray diffraction and transmission electron microscopy measurements (results to be published elsewhere).
At the end of the growth the heterostructures were cooled down to ambient temperature in \SI{5e-3}{\milli\bar} oxygen pressure (further details in Supplementary Figure 1).

\textit{Hall bar geometry fabrication:}
\ce{SrTiO3} substrates were patterned prior to \ce{WO3 / LaAlO3} thin films deposition with standard e-beam lithography followed by the evaporation of an insulating \ce{Al2O3} mask.
The mask was deposited at room temperature by RF sputtering in a \SI{5}{\micro\bar} \ce{Ar} atmosphere, resulting in amorphous alumina.

\textit{Electrical measurements:}
The measurements in \cref{fig:PhaseDiagram_MIT,fig:MagnetoTransport} were carried out in van der Pauw configuration, while for the ones in \cref{fig:MT_temperature,fig:SdH} a Hall bar geometry was used.
In both measurement configurations the metallic interface was directly contacted by ultrasonically wire-bonded \ce{Al}.

\textit{Non-linear Hall effect fits:}
The fits are performed with the least squared method using data in the magnetic field range \SIrange{-4}{4}{\tesla}.
The constraint $1/\rho_0=1/\pL+1/\pH$ is applied to the fitting parameters, and $\rho_0$ is extracted from the $\rho_{xx}(B)$ measurement.
With the assumption $1/\rho_i = n_i e \mu_i$, only three free parameters among {$\rho_i$, $n_i$, $\mu_i$}, with $i={\mathtt{I},\mathtt{II}}$, are varied in the fitting procedure.
} 

\section{Acknowledgments}
We thank P. Zubko for valuable feedback and for performing XRD measurements; Y. M. Blanter and D. J. Groenendijk for fruitful discussions.
This work was supported by
The Netherlands Organisation for Scientific Research (NWO/OCW) as part of the Frontiers of Nanoscience program (NanoFront),
the Dutch Foundation for Fundamental Research on Matter (FOM),
the European Research Council under the European Union's H2020 programme/ ERC GrantAgreement n. [677458]
and the Cornell Center for Materials Research with funding from the NSF MRSEC program (DMR-1120296).
The FEI Titan Themis 300 TEM was acquired through NSF-MRI-1429155, with additional support from Cornell University, the Weill Institute and the Kavli Institute at Cornell.
A. F. thanks TU Delft and Kavli Institute for the access to Computing Center resources, and computational support from the CRS4 Computing Center (Piscina Manna, Pula, Italy).

\bibliography{Biblio}

\newif\ifpaper

\papertrue

\ifpaper
\onecolumngrid
\appendix
\newpage
\noindent\rule{1\columnwidth}{1pt}
\section{\huge \texttt{Supplementary Information}}
\noindent\rule{1\columnwidth}{1pt}
\else
\documentclass[aps,preprint,superscriptaddress,groupedaddress]{revtex4}
\usepackage{graphicx} 	
\usepackage{dcolumn} 	
\usepackage{bm}   		
\usepackage{amssymb}   	
\usepackage[separate-uncertainty=true]{siunitx}		
\usepackage[svgnames]{xcolor}	
\usepackage[version=3]{mhchem}	

\hyphenation{ALPGEN}
\hyphenation{EVTGEN}
\hyphenation{PYTHIA}

\usepackage[caption=false]{subfig}
\usepackage[hidelinks]{hyperref}
\usepackage{cleveref}
\begin{document}
\title{\texttt{Supplementary Information}\\
High mobility electron systems at multipolar oxide interfaces}
\fi

\renewcommand{\figurename}{Supplementary Figure}
\setcounter{figure}{0}

\begin{figure*}[h]
	\includegraphics[page=6,width=.7\linewidth]{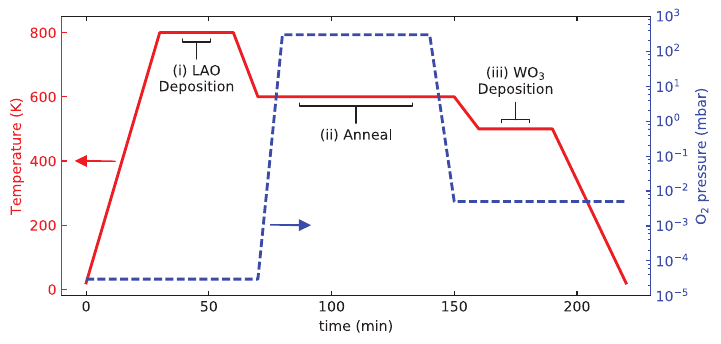}
	
	\caption{\textbf{Growth conditions schematics.}
		Cycle of sample temperature and oxygen pressure used for the growth of all the $(m,n)$ heterostructures:
		(i) heating to \SI{800}{\celsius} in \SI{5e-4}{\milli\bar} followed by possible \ce{LaAlO3} deposition,
		(ii) post-growth anneal step for \SI{1}{\hour} at \SI{600}{\celsius} and \SI{300}{\milli\bar},
		(iii) cool-down to \SI{500}{\celsius} and \SI{5e-3}{\milli\bar} followed by possible \ce{WO3} deposition,
		final cool-down to room temperature.
		A rate of \SI{20}{\celsius\per\minute} was used for all temperature ramps.
		The $(m,0)$ and $(0,n)$ samples underwent the same $T$, $p$ cycle, where the \ce{LAO} or \ce{WO3} deposition steps were substituted by an idling stage of a few minutes.
	}
	
	\label{fig:GrowthConditions}
	
\end{figure*}

\begin{figure*}[h]
	\includegraphics[page=7,width=.8\linewidth]{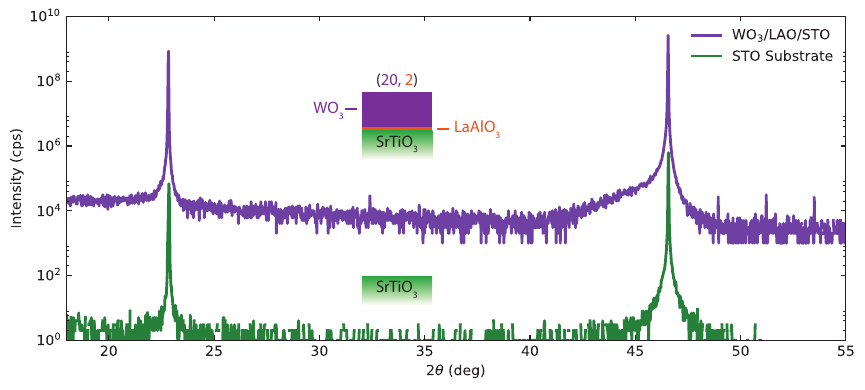}
	
	\caption{\textbf{XRD comparison of an \ce{STO} substrate and a $(20,2)$ \ce{WO3 / LAO / STO} heterostructure.} 
		The $\theta-2\theta$ X-ray diffraction scan around the (001) and (002) substrate peak shows no clear diffraction peak coming from the heterostructure, further confirming the amorphous nature of our \ce{WO3} overlayer.
	}
	
	\label{fig:XRD}
	
\end{figure*}

\newpage
\begin{figure*}[h]
	\includegraphics[page=8,width=.7\linewidth]{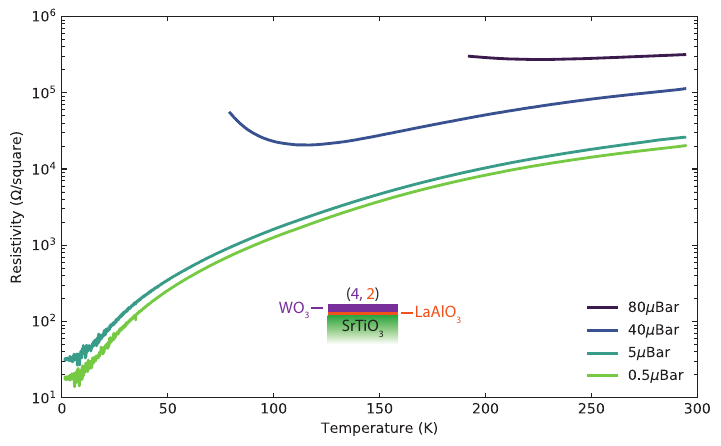}
	
	\caption{\textbf{Transport of a $(4,2)$ \ce{WO3 / LAO / STO} heterostructures where \ce{WO3} is grown in different oxygen pressures.}
		We measure the metallic trend discussed in the paper only when the \ce{WO3} overlayer is grown in a low oxygen pressure ($p_{\ce{O2}} <\SI{40}{\micro\bar}$).
		This shows the important role of defect formation, such as oxygen vacancies, in the \ce{WO3} overlayer in order to obtain the enhanced 2DES at \ce{WO3 / LAO / STO} heterostructures.
	}
	
	\label{fig:pO2}
	
\end{figure*}

\begin{figure*}[h]
	\includegraphics[page=9,width=1\linewidth]{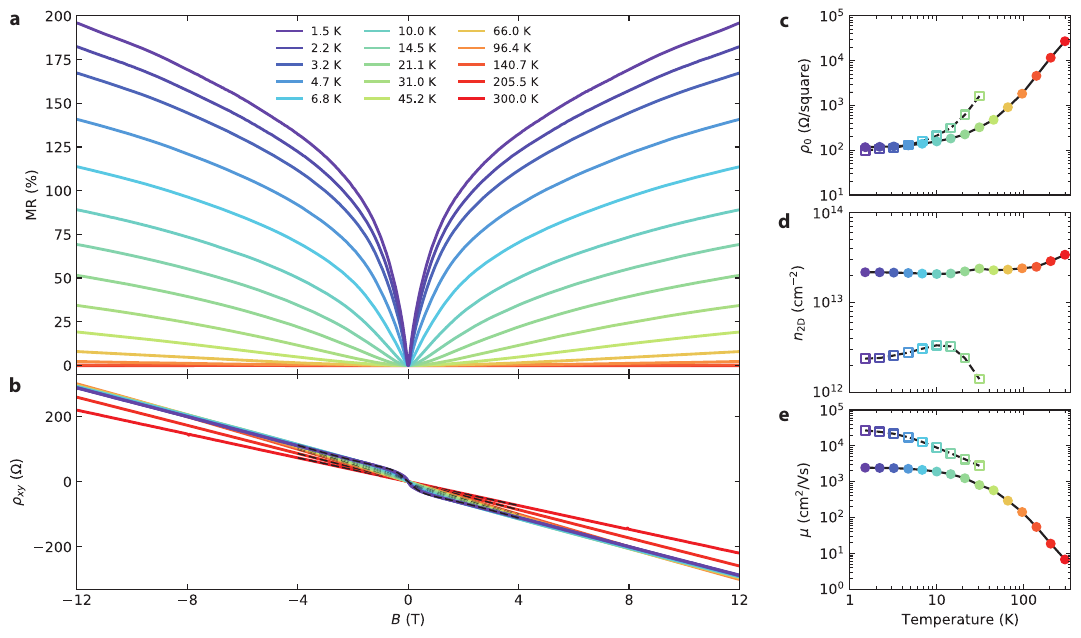}

	\subfloat{\label{fig:MR_T}}
	\subfloat{\label{fig:HE_T}}
	\subfloat{\label{fig:p0_T}}
	\subfloat{\label{fig:n_T}}
	\subfloat{\label{fig:mu_T}}

	\caption{\textbf{Temperature dependent magnetotransport of a $(4,2)$ \ce{WO3 / LAO / STO} heterostructure.}
	\protect\subref{fig:MR_T} Magnetoresistance and
	\protect\subref{fig:HE_T} Hall effect measured at different temperatures in a Hall bar configuration.
	Fits of the Hall effect curves allow to extract
	\protect\subref{fig:p0_T} sheet resistance,
	\protect\subref{fig:n_T} carrier density and
	\protect\subref{fig:mu_T} mobility of the multiple conduction channels.
	}
	
	\label{fig:Magnetotransport_T}
	
\end{figure*}

\begin{figure*}[h]
	\includegraphics[page=10,width=1\linewidth]{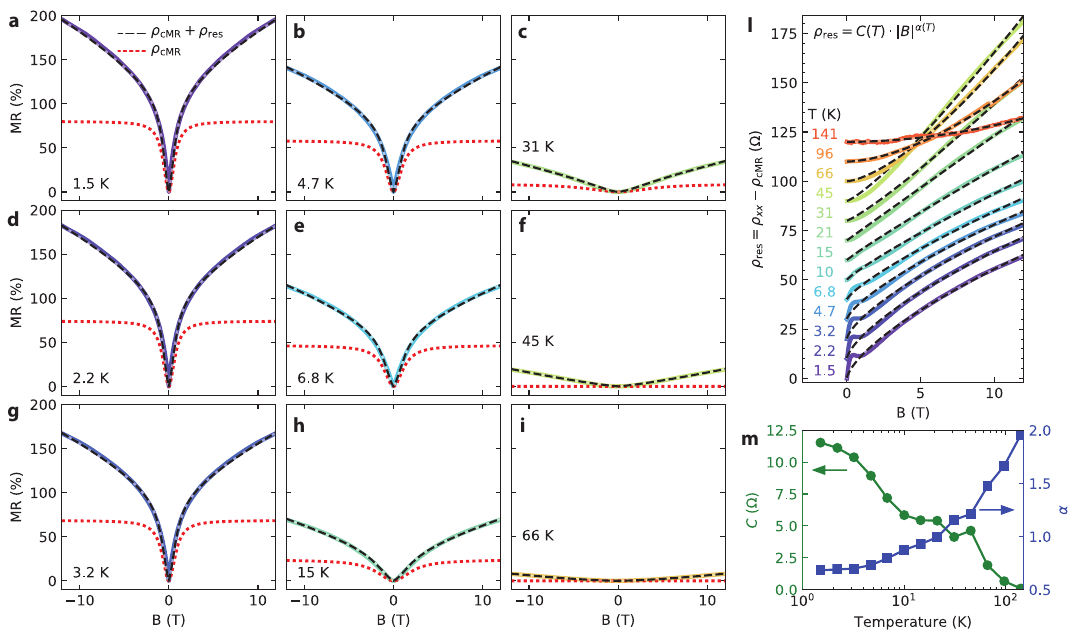}
	
	\subfloat{\label{fig:MR_1K}}
	\subfloat{\label{fig:MR_2K}}
	\subfloat{\label{fig:MR_3K}}
	\subfloat{\label{fig:MR_4K}}
	\subfloat{\label{fig:MR_6K}}
	\subfloat{\label{fig:MR_15K}}
	\subfloat{\label{fig:MR_31K}}
	\subfloat{\label{fig:MR_45K}}
	\subfloat{\label{fig:MR_66K}}
	\subfloat{\label{fig:MR_residual}}
	\subfloat{\label{fig:MR_residual_params}}
	
	\caption{\textbf{Phenomenological magnetoresistance fits.}
		\protect\subref{fig:MR_1K} to \protect\subref{fig:MR_66K} Experimental magnetoresistance (solid line) at different temperatures, classical magnetoresistance $\rho_\mathrm{cMR}$ (red dotted line) calculated with the parameters extracted from the Hall effect and total magnetoresistance (black dashed line), obtained summing the residual magnetoresistance fit $\rho_\mathrm{res}$ with $\rho_\mathrm{cMR}$.
		\protect\subref{fig:MR_residual} Residual magnetoresistance curves (offset for clarity) are fitted with a power law function (dashed black lines) and the extracted parameters are in
		\protect\subref{fig:MR_residual_params}.
		The fits, performed in the \SIrange{3}{12}{\tesla} range, well represent the high-field part of the experimental curves.
		A decreasing trend as a function of temperature is found for the amplitude factor $C(T)$, consistent with the decreasing magnetoresistance effect.
		The exponent $\alpha(T)$, instead, grows with increasing temperature and passes from $\alpha<1$ ($\rho_\mathrm{res}$ curve concave down) to $\alpha>$ ($\rho_\mathrm{res}$ concave up) at about \SI{30}{\kelvin}.
		We note that this temperature corresponds to the crossover from two to one conduction channels observed in the Hall measurements.
		The power-law dependence of this residual magnetoresistance might be related to spatial inhomogeneities of {$n_\mathrm{2D}$, $\mu$} or disorder.
	}
	
	\label{fig:cMR}
	
\end{figure*}

\begin{figure*}[h]
	\includegraphics[page=11,width=.7\linewidth]{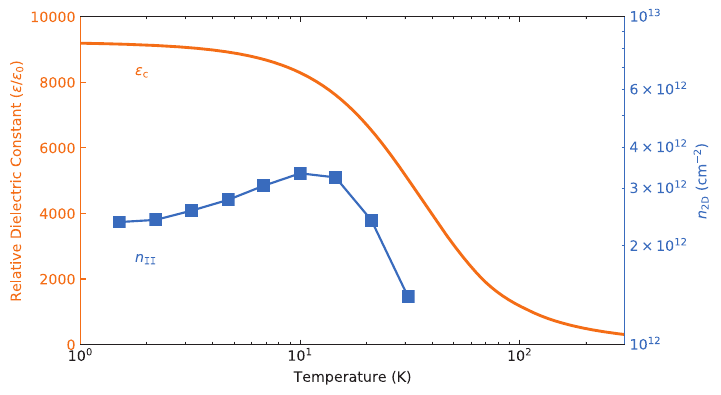}
	
	\caption{\textbf{Comparison of STO dielectric constant with high-mobility carrier density.}
	Temperature dependence of the $c$-axis relative dielectric constant of STO (orange) from [Sakudo T. et al., PRL (1971)] and carrier density of the high mobility channel of the $(4,2)$ \ce{WO3 / LAO / STO} heterostructure in \cref{fig:n_T}.
	We note how $\nH$ decreases dramatically and then disappears in the same temperature range where the STO dielectric constant drops, suggesting the loss of high-mobility carriers can be related to a change in the STO dielectric environment.
	}

	\label{fig:STO_dielectric}
	
\end{figure*}

\end{document}

\end{document}